\documentclass[a4paper,fleqn,usenatbib]{mnras}

\usepackage{newtxtext,newtxmath}

\usepackage[T1]{fontenc}
\usepackage{ae,aecompl}
\usepackage{placeins}

\usepackage{graphicx}	
\usepackage{amsmath}	
\usepackage{amssymb}	
\usepackage{amsfonts}
\usepackage{colors}
\usepackage[normalem]{ulem}
\usepackage[usenames,dvipsnames]{color}

\title[Massive data compression for parameter-dependent covariance matrices]{Massive data compression for parameter-dependent covariance matrices}
\author[Heavens, A. F., Sellentin, E., de Mijolla, D., Vianello, A.]{Alan F. Heavens$^{1}$\thanks{a.heavens@imperial.ac.uk}, Elena Sellentin$^{1,2}$, Damien de Mijolla$^1$, Alvise Vianello$^1$\\
$^1$Imperial Centre for Inference and Cosmology (ICIC), Astrophysics, Imperial College, Blackett Laboratory, Prince Consort Road, London SW7 2AZ, U.K.\\
$^2$D\' epartement de Physique Th\' eorique, Universit\' e de Gen\` eve, Quai Ernest-Ansermet 24, CH-1211 Gen\` eve, Switzerland}

\newcommand{\be}{\begin{equation}}  \newcommand{\ee}{\end{equation}}
  \newcommand{\ba}{\begin{eqnarray}}
\newcommand{\ea}{\end{eqnarray}}

\newcommand{\B}{{\sf B}}
\newcommand{\C}{{\sf C}}
\newcommand{\sS}{{\sf S}}
\newcommand{\F}{{\sf F}}
\newcommand{\I}{{\sf I}}

\newcommand{\bb}{{\bf b}}   
  
  \newcommand{\br}{{\bf r}}

\newcommand{\bx}{{\bf x}}  
\newcommand{\by}{{\bf y}}

\newcommand{\bbmu}{\mbox{\boldmath $\mu$}}

\def\gs{\mathrel{\raise1.16pt\hbox{$>$}\kern-7.0pt %
\lower3.06pt\hbox{{$\scriptstyle \sim$}}}}         %
\def\ls{\mathrel{\raise1.16pt\hbox{$<$}\kern-7.0pt %
\lower3.06pt\hbox{{$\scriptstyle \sim$}}}}         %

\begin{document}

\voffset=-0.25in 
\maketitle

\begin{abstract}
We show how the massive data compression algorithm MOPED can be used to reduce, by orders of magnitude, the number of simulated datasets that are required to estimate the covariance matrix required for the analysis of gaussian-distributed data.  This is relevant when the covariance matrix cannot be calculated directly. The compression is especially valuable when the covariance matrix varies with the model parameters.  In this case, it may be prohibitively expensive to run enough simulations to estimate the full covariance matrix throughout the parameter space.  This compression may be particularly valuable for the next-generation of weak lensing surveys, such as proposed for Euclid and LSST, for which the number of summary data (such as band power or shear correlation estimates) is very large, $\sim 10^4$, due to the large number of tomographic redshift bins that the data will be divided into.  In the pessimistic case where the covariance matrix is estimated separately for all points in an MCMC analysis, this may require an unfeasible $10^9$ simulations.  We show here that MOPED can reduce this number by a factor of 1000, or a factor of $\sim 10^6$ if some regularity in the covariance matrix is assumed, reducing the number of simulations required to a manageable $10^3$, making an otherwise intractable analysis feasible.
\end{abstract}


\section{Introduction}

Many problems concern data that are gaussian-distributed, either as a result of some underlying physical process, or by virtue of the central limit theorem.  The sampling distribution then depends only on the mean and the covariance matrix of the data, and inference of model parameters then follows with the use of a likelihood that is a multivariate gaussian function of the data.  One challenge that can be considerable is if the covariance matrix cannot be calculated readily, and the experiment has to be simulated and the covariance matrix estimated from the simulated data.  In principle this is not difficult, but it can be expensive to do, since at least $p+3$ simulations are required, where $p$ is the number of data.  If the number of simulations is less than this, the expectation of the precision matrix (the inverse of the covariance matrix) diverges.  Ideally one would like many more than $p+3$, in order for the estimated covariance matrix to be precise. Furthermore, if the covariance matrix depends on the model parameters, then it may be a severe challenge: for Bayesian inference using, for example Monte Carlo Markov Chains, the covariance matrix might in the worst case be estimated at each point in parameter space that is sampled. 

If it is impractical to perform so many simulations, then some savings may be made by regularising the behaviour of the covariance matrix, but in addition we can markedly improve the situation by reducing the number of data points $p$, in some cases by orders of magnitude.  In the same spirit, \cite{AS} proposed a more modest level of linear compression of COSEBI statistics. In general, this will lose information, but we previously published an algorithm MOPED\footnote{Massively Optimised Parameter Estimation and Data compression} \citep{HJL} which can massively reduce the number of data points, without losing information, in the sense that the Fisher matrix is unchanged by the data compression, subject to certain conditions. The MOPED algorithm reduces the size of the dataset from $p$ to $m$, where $m$ is the number of parameters in the model, and this can be a dramatic reduction in the dataset size with little or no loss of information.  It has been successfully applied to determine the star formation history of galaxies \citep{RJH,Heavens2004,Panter}, and investigated for data compression in the Cosmic Microwave Background \citep{GH,ZD} and in gravitational waves \citep{Graff}.

MOPED is therefore an interesting candidate to tackle the issue of experiments with large datasets, relatively few model parameters, and covariance matrices that need to be simulated.  In this paper, we explore how effective MOPED can be in such situations, finding that it can reduce enormously the computational requirements to analyse such experiments, at the expense of a small increase in the parameter errors compared with the ideal, but unattainable, analysis.

The second element in this paper is that when the covariance matrix is estimated, then the true covariance matrix needs to be marginalised over, as shown by \citet{SH}, leading to a modified $t$-distribution.  This leads to a modification of the credible regions, increasing them at low credibility levels but maintaining a compact core.  The situation may be more complicated if one has some prior knowledge of the covariance matrix, or there is one part of it that is known.  This subject has become very topical in the light of the expected dataset size of future cosmology surveys such as Euclid and the Large Synoptic Survey Telescope (LSST), and as a result, much attention is being devoted to this issue. As one application, it is expected that next generation photometric surveys will be split into $\sim10$ tomographic bins of redshift, so with $\sim 25$ band-powers in frequency (or separations, if configuration-space statistics such as correlation functions are used), then the total number of summary data, including auto- and cross-correlations, and $E$ and $B$ (or $\xi_+$ and $\xi_-$), is $\sim 6 \times 10^3$, or higher if one also investigates $E$-$B$ correlations to test isotropy.  

See \citet{SH17} and \citet{Blot,Dodelson} and \citet{Percival,TaylorJoachimi} for assessments of the increase in errors due to uncertainties in the covariance matrix,  and for further discussion, see \citet{Joachimi,FriedrichEifler} and \citet{Padmanabhan,Petri,Pope}.  

\section{The MOPED algorithm}

Here we review and extend the MOPED algorithm as originally presented in \citet{HJL}.  MOPED forms linear combinations of the data $\bx$ (which has length $p$), using a set of MOPED vectors $\bb_\alpha$, where $\alpha=1\ldots m$, and $m$ is the number of parameters, each of which is contained in an ordered list represented by a vector $\btheta$. They compress the data to a set of MOPED coefficients
\be
y_\alpha = \bb_\alpha^T\bx.
\label{ybx}
\ee 
The MOPED vectors are chosen in sequence, according to the following algorithm:  the first is the linear combination that minimises the expected conditional error on parameter $\btheta_1$.  i.e. it maximises the matrix element $\F^y_{11}$, where $\F^y_{\alpha\beta} = -\langle \partial^2\ln L^y/\partial\theta_\alpha\partial\theta_\beta\rangle$ is the Fisher matrix for the $\by$ dataset, and $L^y = p(\by|\btheta)$ is the likelihood of the compressed data.  Subsequent $\bb_\alpha$ vectors are chosen to maximise $\F^y_{\alpha\alpha}$ ($\alpha>1$), subject to the $\bb$ vectors being orthogonal to the previous vectors, in the specific sense that the $y_\alpha$ are uncorrelated.  This requires $\bb_\alpha^T \C \bb_\beta = \delta_{\alpha\beta}$, if we also normalise the MOPED coefficients to unit variance.

For gaussian data, the Fisher matrices ($\F^y$ and the analogue $\F^x$ for the original dataset) are computed from \citep{TTH}
\be
\F_{\alpha\beta} = \frac{1}{2}{\rm
Tr}\left[\C^{-1}\C_{,\alpha}\C^{-1} \C_{,\beta}
+ \C^{-1}(\bbmu_{,\alpha}\bbmu_{,\beta}^T + \bbmu_{,\beta}\bbmu_{,\alpha}^T)
\right].
\label{FisherFull}
\ee
where $\bbmu \equiv \langle \bx(\btheta)\rangle$ or $\langle \by(\btheta)\rangle$ are length $p$ or length $m$ expected data vectors for the full or compressed datasets respectively.  $\C$ is the $p \times p$ or $m \times m$ covariance matrix of the data, and a comma indicates a partial derivative with respect to the labelled parameter.   Generally if there is no superscript on $\C$, it will refer to the original data vector $\bx$, but we will identify $\C$ with a superscript $x$ or $y$ if extra clarity is required.  We assume that $\bbmu$ and its derivatives can be computed via theoretical or computational methods.

In its previous applications, MOPED has made the assumption that the covariance matrix is independent of the parameters.  This assumption is relaxed in this paper, and we can properly account for the parameter dependence. MOPED also requires a fiducial set of parameters to be chosen, since the Fisher matrix depends on derivatives of $\bbmu$ as well as the covariance matrix.  The solutions for the optimised weighting vectors in eq.(\ref{ybx}) are
\be
\bb_1 = \frac{\C^{-1} \bbmu_{,1}}{\sqrt{\bbmu_{,1}^T\C^{-1}\bbmu_{,1}}}
\label{Evector1}
\ee
and
\be
\bb_\alpha = \frac{\C^{-1}\bbmu_{,\alpha} - \sum_{\beta=1}^{\alpha-1}(\bbmu_{,\alpha}^T 
\bb_\beta)\bb_\beta}{
\sqrt{\bbmu^T_{,\alpha} \C^{-1} \bbmu_{,\alpha} - \sum_{\beta=1}^{\alpha-1}
(\bbmu_{,\alpha}^T \bb_\beta)^2}}\qquad 1<\alpha\le m,
\label{bbm}
\ee
where $\C$ and $\bbmu_{,\alpha}$ are evaluated at the fiducial parameter set.

It can be shown \citep{HJL} that if the fiducial parameters coincide with the true parameters, and the Fisher matrix is dominated by the second term of eq.(\ref{FisherFull}), then the compression is locally lossless, defined by $\F^x = \F^y$.  In the case that the covariance matrix does not depend on the parameters, the covariance matrix of the compressed data is by construction very simple everywhere.  If we define $\B$ to be a $p \times m$ matrix of which the columns are the $\bb$ vectors, then the compressed data vector is $\by = \B^T\bx$, and from the orthogonality condition of the $\bb$ vectors, 
\be
\C^y = \B^T \C \B = \I_m
\label{Cy}
\ee
i.e. the $m\times m$ identity matrix.  This makes parameter inference with MOPED extremely fast, as the likelihood involves only $O(m)$ operations, rather than the $O(p^3)$ operations for the full dataset, provided that the covariance matrix is independent of the model parameters.  Note that the method is completely general: the data and the model can be anything.  In this paper, we use as an illustrative example the pixellised intensities of a galaxy image as the data vector, and an exponential light profile as the model.    Another example, which we do not explore, is to take as the data vector the estimates of the shear correlation functions in a weak lensing analysis. A specific example of this is the CFHTLenS analysis of \citet{Heymans13}, which had $p=210$ shear correlation measurements and a model with $m=6$ parameters, so the gains would be considerable in this case.  However, since various assumptions in deriving eq.(\ref{Cy}) are violated in practice, in this paper we do not assume that $\C^y$ is the identity matrix, but we estimate it with simulations (see \S \ref{Cy}).  

\section{Parameter inference with estimated covariance matrices}

The fact that the covariance matrix is not known but is estimated changes the likelihood function.  As pointed out by \citet{Kaufman}, even if the estimated covariance matrix is unbiased, its inverse is not, and needs to be multiplied by a factor $\alpha=(N-p-2)/(N-1)$ to make it so, where $p$ is the number of data and $N$ the number of simulations.  This was introduced into astronomy by \citet{Hartlap}.  In fact it is strictly incorrect to retain the gaussian form and use an unbiased inverse covariance matrix; rather one should marginalise over the true covariance matrix, given its estimate.  Since the estimate, which we denote by $\sS$, follows a Wishart distribution, this can be done analytically, and the solution is given by \citet{SH}, yielding a likelihood which is a modified $t$-distribution:
\be
\ln p(\bx | \bbmu, \sS, N) = {\rm const.}-\frac{N}{2} \ln \left[ 1 + \frac{(\bx -\bbmu)^T\sS^{-1} (\bx - \bbmu) }{N-1}\right] .
  \label{tdistrib}
\ee
In the limit $N\gg p$, this approaches the original gaussian distribution, but in general it has a narrower core and wider tails. 

If we form linear combinations of the original data, as here with the MOPED compression, $\by = \B^T\bx$, the estimated covariance matrix of the compressed data $\sS^y$ is also Wishart distributed (with scale matrix $\C^y/n_c$ and degrees of freedom $n_c=N_c-1$, where $N_c$ is the number of simulations of the compressed data).  The same marginalisation then applies, and the likelihood of $\by$   is given by the t-distribution of eq.(\ref{tdistrib}) but with $\bx \rightarrow \by$, $\sS\rightarrow \sS^y$ and $N\rightarrow N_c$, and $\bbmu$ is the expectation value of $\by$.  However, the big advantage of the compression is that the enlargement of the credible regions due to the uncertainty in the covariance matrix is small provided only that $N_c\gg m$, which requires far fewer simulations than when using the full dataset, if $p\gg m$.

\section{Method for a covariance matrix that depends on model parameters}
\label{Cy}

\begin{figure}
\includegraphics[width=8cm, angle=0]{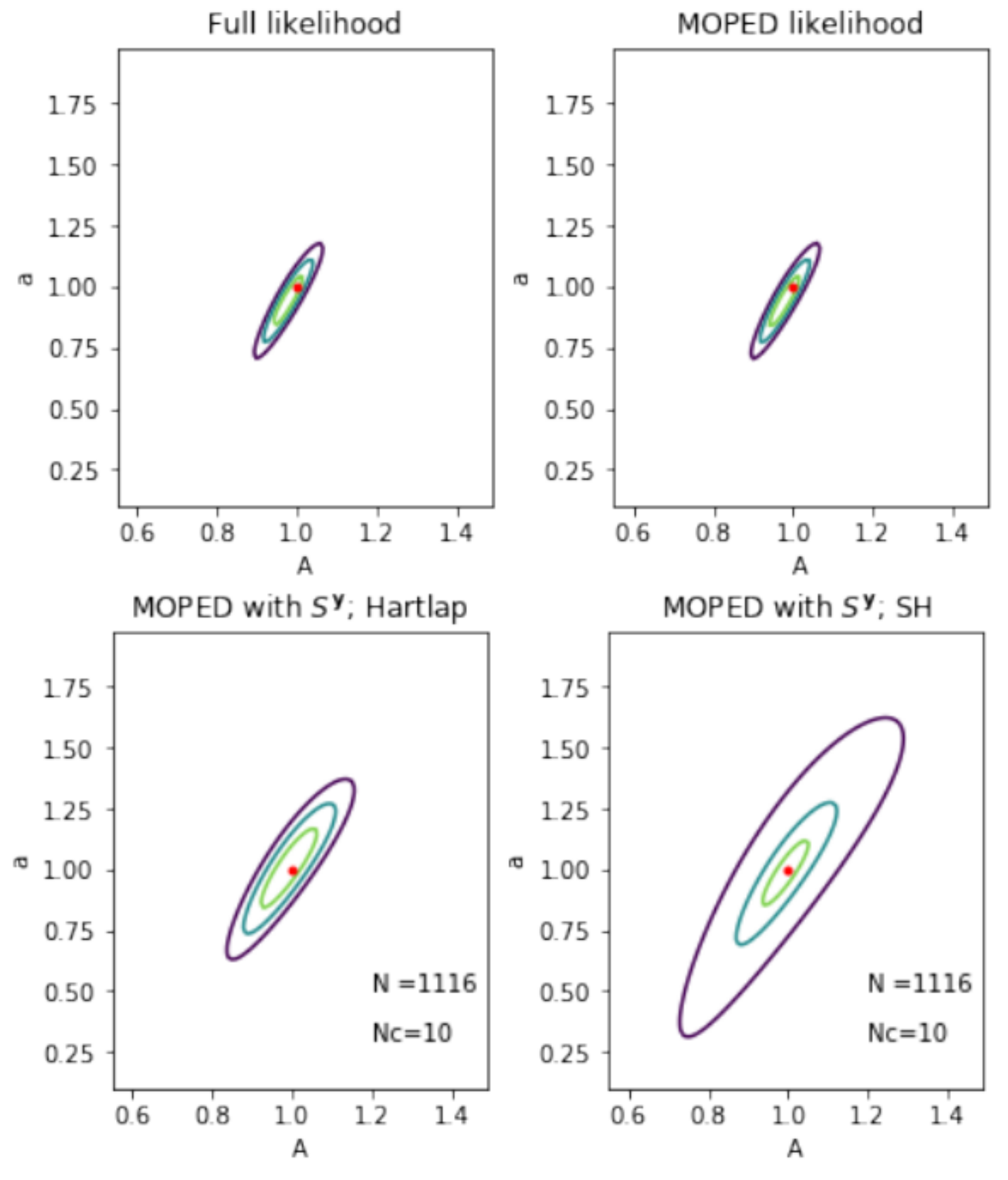}
\caption{Posteriors for the parameters in the model $\mu=A\exp(-a r)$ for 400 pixels and  noise per pixel of 0.1. True values ($A=1,\ a=1$) are marked by the red dot, and contours are at levels $\delta\ln L=-2.3, -6.2, -11.8$, corresponding to $1,2,3\sigma$, 2-parameter credible regions of gaussian likelihoods.  From top left, clockwise: full likelihood with known covariance matrix (best possible case); MOPED compression, using correct full and compressed covariance matrix, and correct fiducial model; compressed analysis using the likelihood of \citet{SH}, with 1116 simulations used to determine the MOPED vectors, and only 10 to estimate the compressed covariance matrix; same, with a gaussian likelihood, using the \citet{Hartlap} scaling, where the inner contour is too large and the outer one is too small. Here we assume that the covariance matrix is parameter-independent, so we estimate it only once.}
\label{Likelihoods}
\end{figure}

\begin{figure}
\includegraphics[width=8cm, angle=0]{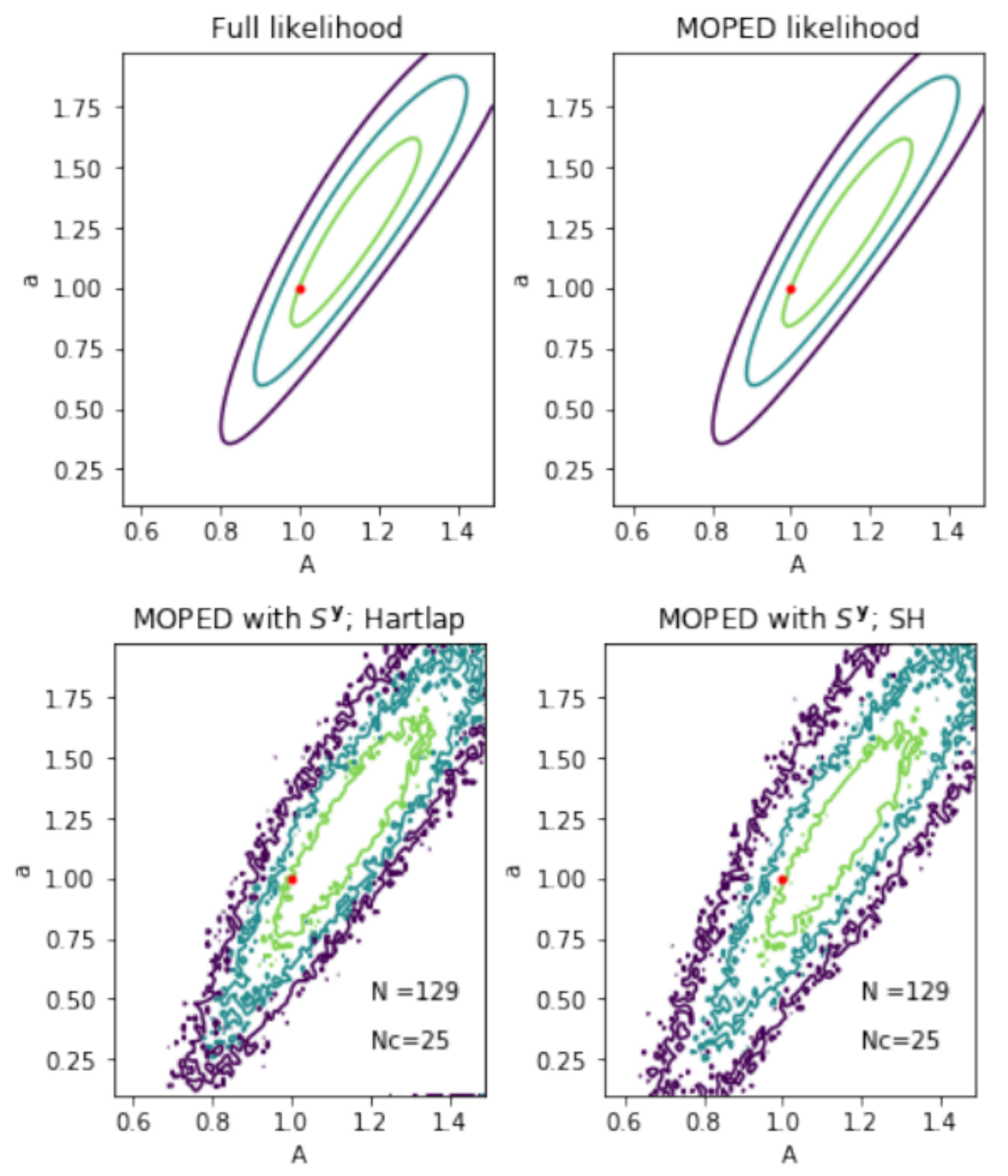}
\caption{Similar to Fig. \ref{Likelihoods}, except that we estimate the compressed covariance matrix separately at each point in the parameter space grid.  This would be required if the covariance matrix varied with the parameters of the model, when brute force estimation of the covariance matrix everywhere might be impractical without the data compression proposed here. There are 25 pixels in this example, the MOPED vectors are determined from 129 simulations, and 25 simulations are used to estimate the compressed covariance matrix at each point. }
\label{NoisyL}
\end{figure}
It is important to realise that we can choose to make any linear compression of the data, whether it is locally lossless or not.  Hence, we can apply a MOPED compression to the data even if the assumptions in its derivation are violated so that the compression is not optimal.   The parameter inference would still be entirely valid; the credible regions would just be larger than they could be.  Past investigations \citep{HJL,GH} have shown that in practical cases, the increase in parameter credible reasons is usually negligibly small.  There is a subtlety in that the inference will be correct provided that the compressed covariance matrix is correct.  In typical MOPED applications, the compressed covariance matrix has been assumed to be fixed at the identity, and this gives very rapid inference. However, it is an approximation if $\C$ depends on parameters. If more accuracy and precision are required, an iterative solution is to find the most probable parameters and then repeat the MOPED data compression with the solution as the fiducial model.  In practical applications we have not found this to be necessary, but strictly we should use the correct covariance matrix appropriate for the position in parameter space.  This is what we do in this paper.

An alternative to assuming that the compressed covariance matrix is the identity is to compute it directly from $\B$ and $\C$, or equivalently to simulate $\bx$ and then form $\by$ by matrix multiplication, and estimate $\sS^y$ from the simulated $\by$ vectors.  This then dispenses with an approximation, but the cost of the matrix operations in normal applications then negates the massive computational speed advantage of MOPED.   However, in this paper we are considering the situation where the time is dominated by the time to estimate the covariance matrix, not to evaluate the likelihood, so the matrix operations used to generate $\by$ come at negligible cost. 

The method we advocate is this: create $N$ simulated datasets, $\bx_{(i)}; \ i=1\ldots N$, for a fiducial set of model parameters, in order to obtain an unbiased estimate $\sS$ for the full (fiducial) covariance matrix:
\be
\sS = \frac{1}{N-1}\sum_{i=1}^N (\bx_{(i)} - \bar\bx)(\bx_{(i)} -\bar\bx)^T.
\label{Chat}
\ee
We then use this to precompute a set of MOPED compression vectors, using equations (\ref{Evector1}) and (\ref{bbm}) but with $\C$ replaced by its estimate $\sS$.  This set will be close to optimal, provided that the chosen fiducial model is correct, and the covariance  matrix $\sS$ has been estimated from sufficiently many simulations to be a good approximation to $\C$.   After this point we keep the MOPED $\bb$ vectors fixed, and do not vary them during the parameter inference phase.  If this preliminary step is already too expensive in terms of computer time, then an alternative approach is to use an approximate covariance matrix, perhaps theoretically generated on the basis of assumptions that do not precisely hold.  The MOPED vectors would not be optimal in this case, but may be close enough that the information loss is small.

When inferring parameters (via say MCMC chains), we again make an estimate of a covariance matrix, but this time we form $\sS^y$ as an estimate for the compressed data covariance matrix $\C^y$, using
\be
\sS^y = \frac{1}{N_c-1} \sum_{i=1}^{N_c} (\by_{(i)} -\bar\by)(\by_{(i)} -\bar\by)^T.
\label{Sy}
\ee
The advantage that we have is that we require only $m+3$ or more simulations, rather than the typically much larger $p+3$.   If the simulations are expensive, this could still be a considerable cost, depending on how much the covariance matrix depends on the parameters, but it may make the analysis feasible when otherwise it might be essentially impossible (if $p\gg m$, as is typical).

\section{Example problem}

Let us illustrate with a simple $m=2$ parameter model, representing a circularly-symmetric image of a galaxy with an exponential surface brightness profile.  Ignoring complications of finite pixel size, the model is that the pixel brightness values are 
\be
\bbmu(\br) = A \exp(-a |\br|)
\ee
where $\br$ is the pixel position vector, of length $n_{\rm pix}$. $A$ and $a$ are the model parameters.  Purely for simplicity in this illustrative example, we assume that the true covariance matrix is proportional to the identity, $\C=\sigma^2\, \I_p$, where $\sigma^2$ is the pixel variance.   In this initial example, we will not vary $\sigma^2$ with the parameters.

\begin{figure}
\includegraphics[width=8cm, angle=0]{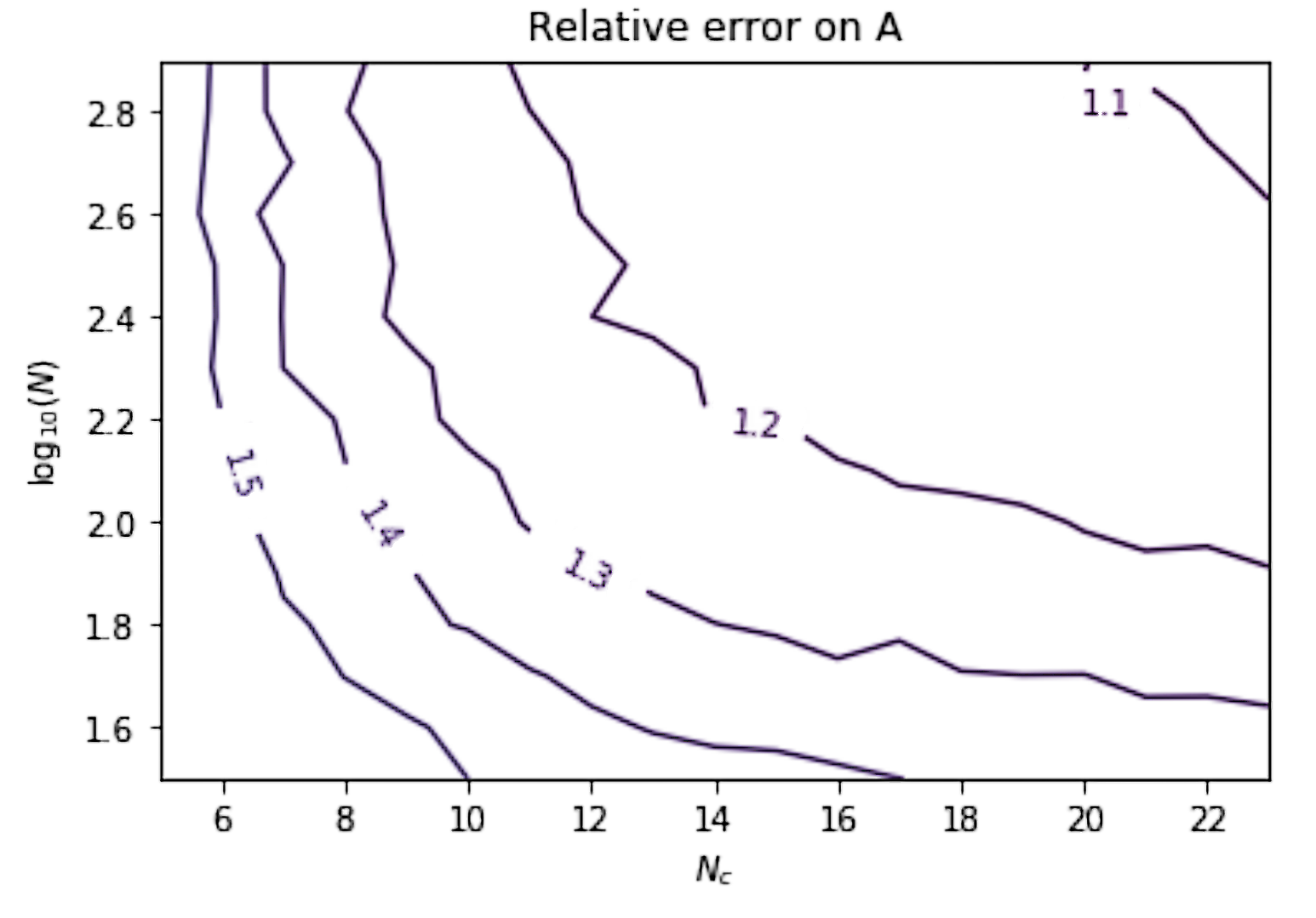}
\caption{Posterior standard deviation for the amplitude parameter $A$, relative to the ideal error when the covariance matrix is known and the full dataset is used. The contour labels refer to the relative increase of the standard deviation. The vertical axis is the number of initial simulations used to estimate the MOPED vectors. The horizontal axis is the number of simulations used to compute the covariance matrix at different points in parameter space. In this case the image is a square of  $p=n^2_{\rm side}=25$ pixels, $\sigma=0.3$ and the plot is averaged over 500 realisations. 
}
\label{sigma1}
\end{figure}

\begin{figure}
\includegraphics[width=8cm, angle=0]{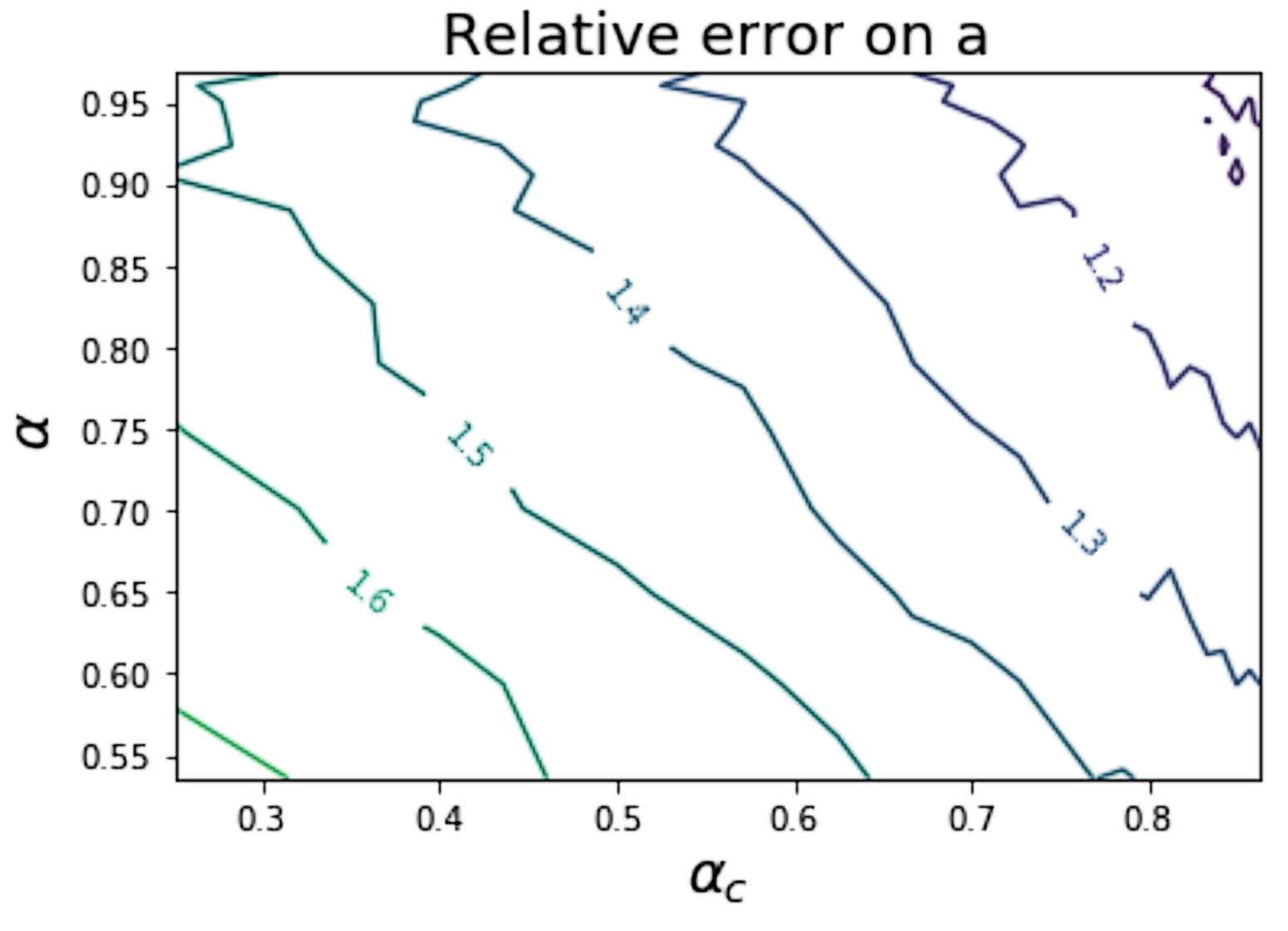}
\caption{As in Fig. \ref{sigma1}, but here for the scale length parameter $a$, and plotted against $\alpha = (N-p-2)/(N-1)$ ($y$ axis) and $\alpha_c = (N_c-m-2)/(N_c-1)$ ($x$ axis), where $m=2$ is the number of compressed data, and $p=25$ is the number of pixels. }
\label{sigma2}
\end{figure}

\begin{figure*}
\includegraphics[width=0.93\textwidth]{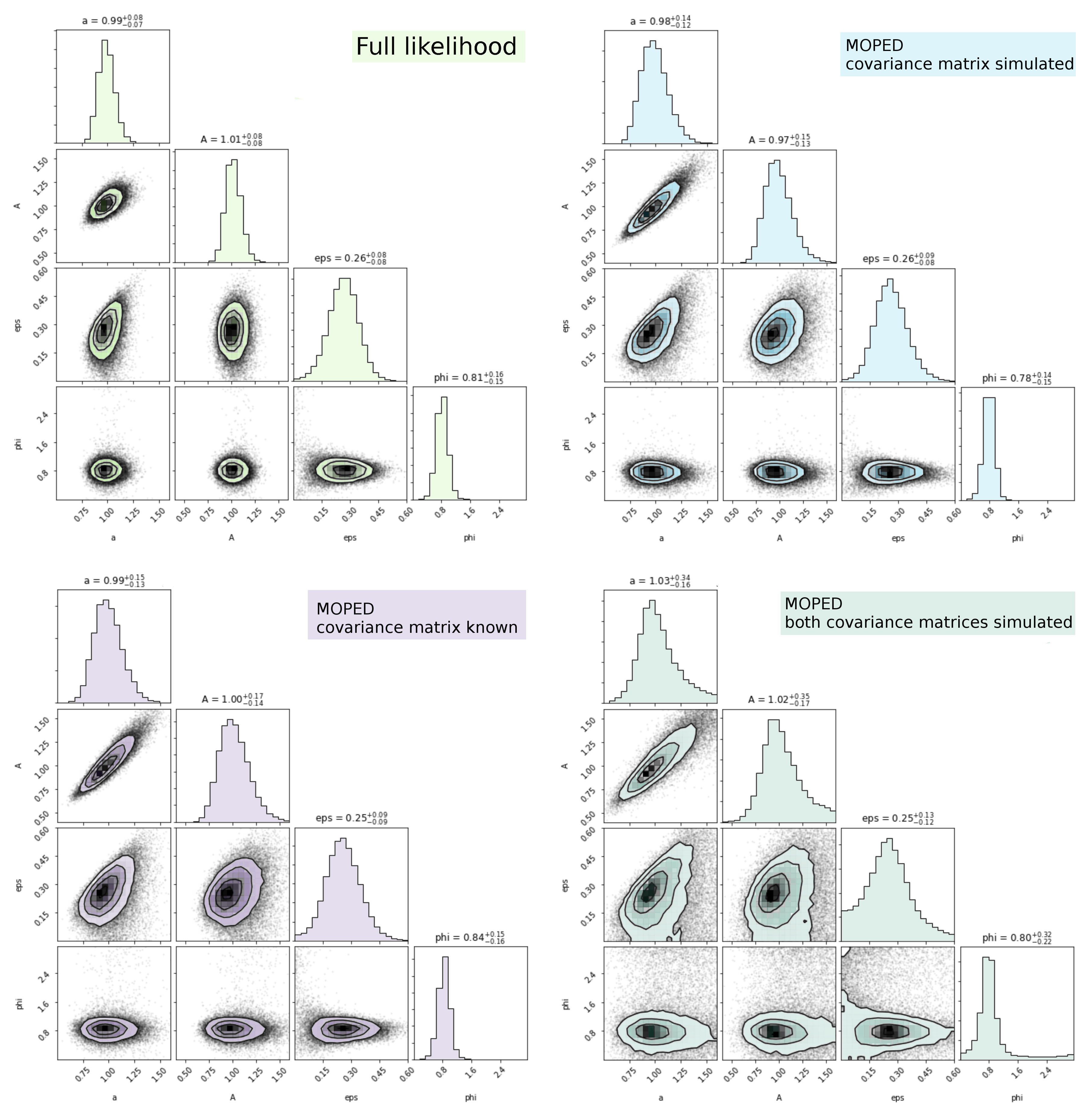}
\caption{\textbf{Top left:}Full likelihood of the 4 parameters $a,A,\epsilon,\phi$ of the model of eq.(\ref{model4}), given a $10\times 10$ galaxy image as the $p=100$ data vector.  10000 points are generated after burn-in, using Hamiltonian Monte Carlo NUTS with \textsc{Stan}. \textbf{Bottom left:} Likelihood of parameters using $m=4$ MOPED compressed data.  MOPED compression vectors have been computed on the basis of the true full covariance matrix. The compressed covariance matrix is assumed to be the identity matrix. Uncertainties are increased since the distribution has been marginalised, and MOPED only guarantees that the distribution is unchanged near the peak. \textbf{Top right:} MOPED compression vectors computed on the basis of an estimated covariance matrix from 1000 simulated datasets. The compressed covariance matrix is assumed to be the identity matrix. In this case, the constraints are essentially as good as if the true covariance matrix was known, since many simulations were used. \textbf{Bottom right:} Now also including parameter-dependence in the covariance matrix with the $4 \times 4$ compressed covariance matrix estimated from 10 simulations at each MCMC point. The credible regions are moderately larger, but the general aim is to make such calculations feasible at all, in cases where it would essentially be impossible to compute the full likelihood. }
\label{Plots}
\end{figure*}

In order to estimate $\C$, we generate $N$ simulated datasets and evaluate $\sS$ via eq. (\ref{Chat}). For a given $N$ we compute the two MOPED vectors using eq. (\ref{Evector1}) and (\ref{bbm}), with $\C$ replaced by $\sS$.  We then generate a test image and compute the likelihood of $A$ and $a$ using the compressed data. In this example, we estimate the compressed covariance only once, from $N_c$ simulations of the full dataset, which are then compressed, and use eq.(\ref{Sy}). 
We then compute the posterior of $A$ and $a$ given the estimated compressed covariance matrix, by analytically marginalising over the unknown covariance matrix, and using the likelihood of \citet{SH}.

In Fig. \ref{Likelihoods} we show contours of the likelihood, for a case when the covariance matrix is independent of parameters, so the compressed covariance matrix is estimated only once. We see that MOPED is very effective when the covariance matrix is known and the fiducial model is the correct one.  In more realistic cases, when the covariance matrices for the full data and for the compressed data have to be estimated, then the compression is not locally lossless except in the limits $N,N_c\rightarrow \infty$.  The large size of the outer contour in the bottom right panel comes from the broad wings of the \citet{SH} likelihood, whereas the contours containing $\sim 68\%$ and $\sim 95\%$ of the posterior are not much larger than in the ideal case.    

If the covariance matrix depends on the parameters of the model, then the analysis is much more challenging.  The covariance matrix may need to be estimated separately each time a new point in parameter space is considered.  Fig.~\ref{NoisyL} is an illustration of this, where we estimate the compressed covariance matrix afresh at every point in the parameter grid.  Since the estimated covariance matrix is a random object, this adds noise to the posterior,  which might benefit from some smoothing. In practice some sort of regularisation procedure would almost certainly be employed for the covariance matrix, which would smooth the contours. 

In Figs. \ref{sigma1} and \ref{sigma2} we show the relative increase in error compared with the ideal case (where we use the true covariance matrix and the full dataset, or indeed the MOPED compressed data assuming the correct fiducial model and covariance matrix; they are essentially identical), as a function of the number of simulations $N$ and $N_c$, or the Hartlap parameters $\alpha=(N-p-2)/(N-1)$ and $\alpha_c=(N_c-m-2)/(N_c-1)$.  To produce these figures we simulate images and compute the marginal credible regions by integration of the 2D posterior, and average over 500 realisations. In Fig. \ref{sigma1} we see that there is little to be gained in increasing $N$ beyond $200$ ($\log_{10}N=2.3$), for this relatively small image of $p=25$ pixels.

In these examples we have chosen datasets of different sizes, $p=400$ and $p=25$ pixels.  The time for inversion of a $p \times p$ matrix scales as $p^3$ (although iterative techniques may be faster), whereas the size of the compressed dataset is $m=2$ in both cases, so the time taken is less dependent on the original dataset size.  The timings would scale as $p^2$, arising both from the time required to generate each sample image used in the estimation of the compressed likelihood, and also from the scalar products that compress the image data.  The generation of the MOPED vectors scales as $p^3$, but this is done only once, and not at each point where the posterior is computed.

\subsection{A more complex model explored with MCMC}

In Fig. \ref{Plots} we show the effect of MOPED compression on a more complex 4-parameter model, which we explore with more typical MCMC techniques.  In this model, the model represents a circular exponential profile disc, seen at an angle, and resulting in a surface brightness distribution (see \cite{HAJ} for more details):
\begin{equation}
\mu(r,\psi | a,\epsilon,\phi,A)=A \exp\left[-{a\,r\sqrt{1+\epsilon^2-2\epsilon \cos 2(\psi-\phi)}}\right]
\label{model4}
\end{equation}
where $A$ is the central surface brightness,
and $a$ is the inverse semi-major axis, $\phi$ its position angle, and $\epsilon$ is the (magnitude of the) ellipticity of the galaxy.  $r$ and $\psi$ are polar coordinates about the centre of the galaxy, whose position is assumed to be known.  Posteriors for the parameters are obtained using \textsc{Stan} \citep{Stan}.  An image is generated with $a=A=\phi=1.0$ and $\epsilon=0.25$, on a $10 \times 10$ grid.  In this case, we make the covariance matrix parameter dependent, assuming white noise, but with a pixel variance that depends on the central amplitude parameter: $\sigma^2=0.01 A$.  

In Fig. \ref{Plots} we plot a comparison of the different possibilities for analysing this dataset. In the top left, we plot the posterior gained from the the full dataset of size $p=100$, using the known covariance matrix. Flat priors on the parameters were assumed. This panel depicts the maximal information content on the parameters to be measured. In the bottom left of Fig. \ref{Plots}, we still assume the correct covariance matrix is known, but we now apply MOPED compression. The MOPED compression vectors are hence determined from the correct covariance matrix, with the fiducial model coinciding with the true model. The covariance for the compressed dataset is then the $4 \times 4$ identity matrix. In the top right, we see the effect of determining the covariance matrix of the full dataset from 1000 simulations, for the purpose of determining the MOPED vectors.  The MOPED compression vectors are therefore not quite optimal, but we still assume the compressed covariance matrix is the identity.  Finally, in the bottom right we show the actual target of MOPED compression in cosmology: parameter dependence in the covariance matrix is now included, and each time a compressed covariance matrix is estimated from 10 simulations only but at each point in the chain. The compressed covariance matrix is then marginalized over using the \cite{SH} likelihood. The likelihood is computed with \textsc{Stan} in a simple hierarchical model, where the covariance matrix is a random object. Note that 10 simulations is not in the asymptotic regime where the compressed covariance matrix is very well determined, so we expect to see a degradation of the errors. We also see that the effect of sampling is to obscure the variability that is apparent in Fig. \ref{NoisyL}.  Note that in this last case, the outer contours are again broadened because of the marginalisation over the true covariance matrix.  The inner contours are only moderately larger than in the other figures, reflecting the small core and broad wings of the Sellentin-Heavens likelihood.  The result of the non-optimal MOPED vectors, and the marginalisation over the compressed covariance matrix are to increase the errors, by approximately 50-100\% in this case. However, the compression has now successfully accomplished the otherwise unfeasible task of computing the parameter-dependence of the (now compressed) covariance matrix at each point of the MCMC chain.

\section{Conclusions}
In this paper, we have considered the relatively common situation of parameter inference from gaussian-distributed data (of length $p$), where the covariance matrix is not directly calculable, but has to be simulated.  In the case where the covariance matrix varies with parameters, this can lead to a requirement for an unfeasibly large number of simulations, especially if the covariance matrix were to be evaluated separately at each sample point in the $m$-dimensional parameter space.  We have shown that this can be speeded up by a very large factor, with little loss of information, by compressing the data using the MOPED algorithm first.  The algorithm proposed is to run a very large number $N \gg p+2$ of simulations with the model parameters kept fixed at some fiducial values, if this is feasible, and to use the resulting estimated covariance matrix for the full dataset to define a set of near-optimal MOPED data compression vectors, which are then kept fixed.  When sampling the parameter space, using MCMC for example, the much smaller compressed covariance matrix may be estimated accurately from far fewer simulations, requiring only $N_c > m+2$, which is typically much less than $p$.

It is clear that there is some trade-off between running many simulations to define the MOPED vectors, and running more simulations during the MCMC phase, but Figs. \ref{sigma1} and \ref{sigma2} indicate that it is likely that the best strategy will be to run $N\gg p$ simulations for a fiducial parameter choice, since an accurate full covariance matrix delivers MOPED vectors that are closer to optimal, and which thus require fewer compressed simulations when the parameter space is sampled.  However, it may be that this reduction in the number of simulations is still inadequate, and there are various possibilities to overcome this.

For the MOPED vectors, it may be adequate to have an approximate full covariance matrix, determined without simulations.  It may not yield optimal compression vectors, but the compression is likely still to be useful. Secondly, to reduce the number of simulations for the compressed stage, one could use some interpolation in the parameter space, estimating the compressed covariance matrix only at a relatively small number of locations.  
  
Emulator-based methods, for example based on a Latin hypercube, may be effective \citep{Heitmann,Heitmann2010}, even with only $\sim 100$ simulations.  Such a scheme was proposed by \cite{MS}, using gaussian processes to interpolate between the covariance matrices. An alternative approximate approach would be to estimate the covariance matrix at a fiducial point, and estimate the generator of the linear part of the variation of the covariance matrix with parameters, and using it to extrapolate to other locations in parameter space \citep{Reischke}.

An additional advantage of this radical data compression is that the central limit theorem may assist in giving the compressed data a near-gaussian sampling distribution, although there is no guarantee that the summary statistics can be grouped into large iid subsets.  Furthermore, it will be far easier to explore numerically the sampling distribution in a small number of dimensions rather than in the original very high dimensional space, to test the Gaussian assumption.

We see clear applications, including, but not limited to, the analysis of weak lensing data from the Euclid and LSST photometric surveys, where the number of summary statistics is expected to be $\sim 10^4$, and the number of cosmological parameters only $\sim 10$, so reductions in the number of simulations by a factor of a thousand is feasible, or by a factor of $10^6$ with emulation techniques as well (see Table 1).

\begin{table*}[tb] 
\begingroup 
\caption{Number of simulations required, for numbers typical of future Euclid or LSST weak lensing surveys, with $p=5000$ summary statistics, $m=6$ cosmological parameters, and an MCMC chain of length $10^5$.  100 emulator points are assumed.}
\label{tab:params}
\footnotesize 

\begin{tabular}{l|c|cll}
\hline
\newdimen\digitwidth 
\setbox0=\hbox{\rm 0}
\digitwidth=\wd0
\catcode`*=\active
\def*{\kern\digitwidth}
\newdimen\signwidth
\setbox0=\hbox{+}
\signwidth=\wd0
\catcode`!=\active
\def!{\kern\signwidth}
\hfill Estimating $C^y$ at: & emulator locations; & each MCMC point. & Comments\cr
\hline
No compression & $10^6$ &  $10^9$ & Estimating $\C^x$ for each MCMC sample is overkill\cr
\hline
MOPED compression, using simulated $\C^x$ & $10^4$ &  $10^6$ & Preferred option \cr
\hline
MOPED compression, using analytic/theoretical $\C^x$ & $10^3$ &  $10^6$ & Sub-optimal, but reduces simulation requirements\cr
\hline
\end{tabular}
\endgroup
\end{table*}

\newpage

\section*{Acknowledgments}
ES is funded by a DAAD research fellowship of the German Academic Exchange Service.



\bsp	
\label{lastpage}
\end{document}